\documentclass[conference]{IEEEtran}
\IEEEoverridecommandlockouts
\usepackage{cite}
\usepackage{amsmath,amssymb,amsfonts}
\usepackage{algorithmic}
\usepackage{textcomp}
\usepackage{graphicx}
\usepackage{float}
\usepackage{xcolor}
\renewcommand{\baselinestretch}{0.95}
\usepackage[hyphens]{url}
\usepackage{hyperref}
\hypersetup{colorlinks=true,breaklinks=true}
\def\BibTeX{{\rm B\kern-.05em{\sc i\kern-.025em b}\kern-.08em
    T\kern-.1667em\lower.7ex\hbox{E}\kern-.125emX}}
\begin{document}

\title{Documentation based Semantic-Aware Log Parsing\\}

\author{\IEEEauthorblockN{Lei Yu, Tian Wu, Jiaqi Li, Patrick Chan, Hong Min, Fanjing Meng}
\IEEEauthorblockA{\textit{IBM Research} \\
Lei.Yu1@ibm.com, tianwu@cn.ibm.com, ljqqi@cn.ibm.com, chanyuk@us.ibm.com, 	hongmin@us.ibm.com, mengfj@cn.ibm.com}
}

\maketitle

\begin{abstract}
With the recent advances of deep learning techniques, there are rapidly growing interests in applying machine learning to log data. As a fundamental part of log analytics, accurate log parsing that transforms raw logs to structured events is critical for subsequent machine learning and data mining tasks. Previous approaches either analyze the source code for parsing or are data-driven such as text clustering. They largely neglect to exploit another widely available and valuable resource, software documentation that provides detailed explanations for the messages, to improve accuracy. In this paper, we propose an approach and system framework to use documentation knowledge for log parsing. With parameter value identification, it not only can improve the parsing accuracy for documented messages but also for undocumented messages. In addition, it can discover the linkages between event templates that are established by sharing parameters and indicate the correlation of the event context. %Our demonstration video can be found here: \url{https://youtu.be/_h-EujDXlqU} 

\end{abstract}


\section{Introduction}
System logs have been widely used in system management and IT operational analytics, because they are universally available in almost all software and systems while also being the primary data resource that records run-time system states and events. Log analytics applies data mining and machine learning learning techniques on log data to model system behaviors and events~\cite{Zong2015, Wu2017a, Yu2016a}, detect system anomalies~\cite{Nandi2016, Fu2009, Du2017a} and security breach~\cite{Shen2018}, and program diagnose~\cite{Xu2009,Shah2017}.

In general, log message is unstructured text produced by logging statements in the software source code. It often consists of invariant keyword part and variable part. The invariant part, that is often referred to as template, represents a message type or an event type, since it remains the same for every event occurrence. The variable part contains the values of specific parameters for the corresponding event, which records the run-time information and can change among different executions and event occurrences. Therefore, for log analytics, log parsing is usually the very first step, which transforms the unstructured raw log data to a sequence of templates through template mining. A sequence of templates represents a sequence of events, since it is tightly stitched together with execution flows. Therefore, most log analytics tasks are based on the sequence of templates to discover the control flow of the system~\cite{Nandi2016} and predict system anomaly~\cite{Fu2009}.

Template mining have been widely studied in recent years. A source code based method~\cite{Xu2009} uses regular expressions automatically generated from the source code for template mining. Most works assume no access to the source code and propose data-drive approaches. For instance, LKE~\cite{Fu2009}, IPLoM~\cite{5936060} and multi-model algorithm~\cite{Nandi2016} use data mining and text clustering techniques. Spell~\cite{Du2017b} and Drain~\cite{He2017} are two online log parsers that use prefix tree to find templates. Data-driven approaches have gained in popularity because the source code are often inaccessible. However, their accuracy is highly affected by the complexity of message text structures, noises and clustering algorithms. The accuracy of mining templates is important, since it decides the accuracy of structured events discovered and the performance of subsequent event mining and system modeling tasks with machine learning techniques.  

In this work, we propose to exploit a widely available but long-neglected resource, software documentation, to improve the performance of log parsing.
Many commercial software products have detailed documentation of log messages such as Oracle StorageTek \cite{oraclestorage},IBM MVS\cite{ibmmvs} and Cisco Security Appliance System\cite{cisco} from different software vendors. These documentations usually define each message with a unique id, a message text that consists of keywords, parameters, and the fields with annotations (in different formats) indicating they are optional or single-select. Besides that, today a growing number of software packages like Kubernetes~\cite{Kubernetes} and MongoDB~\cite{MongoDB} support structured logging in the format of json that is also well documented. The documentation offers rich information of messages that can be used to improve accuracy and efficiency of log parsing even when the source code is inaccessible. 

Our documentation based approach derives the syntax tree of log messages from documentation, and effectively utilizes it for parsing both documented and undocumented log messages. The syntax tree allows more accurate parsing of well documented messages. For undocumented messages, we reuse the parameter values extracted from documented messages to identify the parameter parts in the messages and annotate them with the names given by the documentation. Such parameter name annotation effectively improves the accuracy of template mining. 
With explicit annotation of parameters in a message text, our approach can differentiate the templates with highly similar invariant keywords but distinct parameters semantically, which indicate different events. Previous data-driven approaches parse log messages without being aware of semantics of variable parts. They largely treat the variable part in all messages as the same, which creates confusions that regard different events as the same event. In addition, our approach can discover the semantic correlation between templates. The parameter name annotation can be used to identify if two templates have the same parameter. Sharing the same parameter indicate that two events happen in the same context, which is important information for problem diagnose and system modeling. Since existing log parsing techniques are not able to identify the parameter semantics,they cannot identify the linkage between two templates established by the sharing parameters.

Our demonstration shows the workflow of our documentation based approach, and its effectiveness for parsing of real log messages that are partially documented.    

\section{Semantic-Aware Log Parsing with Software Documentation}\label{sec:approach}

\subsection{Log Message Syntax Tree}
The software documentation may describe the text formats of log messages in many different ways, varying among products, components, and software vendors. Developing a message parser for each type of documentations increases development cost and limits the scalability. To address this problem, we introduce an intermediate representation of log message formats, log message syntax tree, to achieve broad expressiveness of log message formats, while allowing a uniform message parsing for logs from different software products.

We define log message syntax tree with five types of node elements. Two basic types are 
\begin{itemize}
    \item \emph{keyword}:  representing the invariant single-word text in the log message.
    \item \emph{parameter}: representing a named placeholder for the variant part in the log message, carrying the run-time information in every execution.
\end{itemize}
The other three are composite types, which can be composed of any type of elements:  
\begin{itemize}
    \item \emph{sequential}: representing a sequential combination of elements.
The sequential node's children are ordered, representing each element in the sequence. 
    \item \emph{optional}:  representing a special sequential type, of which the element sequence can be either included or excluded from the log message body.
    \item \emph{single-select}: representing a list of the elements of that only one element is included in the log message body.
\end{itemize}
All the leaf nodes in the log message syntax tree are the basic types, and the parent nodes are composite types. 

\begin{figure*}[!htbp] 
\centerline{\includegraphics[height=1in, width=5in]{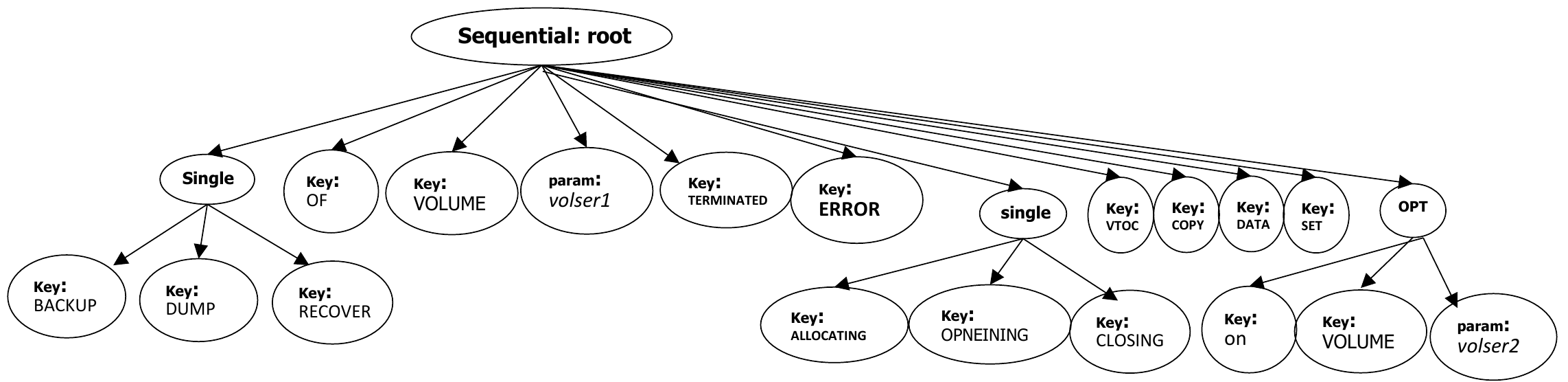}}
\caption{The syntax tree of message ARC0704E}
\label{fig1}
\vspace{-10pt}
\end{figure*}

\noindent\textbf{Example.}
We use a message with identifier ARC0704E from IBM z/OS MVS(Multiple Virtual Storage) System as an example. This message indicates the error on volume operations. Although it has a unique ID, its message text has optional keywords and single-select field, which means that the message instances of ARC0704E can have multiple templates. Its instances can be:
\begin{itemize}\small
\item
example 1: ARC0704E RECOVER OF VOLUME vol1 TERMINATED, ERROR ALLOCATING VTOC COPY DATA SET ON VOLUME vol2
\item
example 2:ARC0704E BACKUP OF VOLUME vol1 TERMINATED, ERROR OPENING VTOC COPY DATA SET
\end{itemize}
where RECOVER, BACKUP are chosen from a list of keywords,\emph{ON VOLUME vol2} is an optional string in the message, and \emph{vol1} \& \emph{vol2} are parameter values.
The syntax tree for message ARC0704E derived from documentation is shown in Figure \ref{fig1}.

The definition of log message syntax tree can cover almost every different format of log messages we've seen due to its rich expressiveness. With the syntax tree, a log message instance can be effectively parsed by depth-first search (DFS), and the parameter values can be easily extracted (Section .

\noindent\textbf{Deriving Message Syntax Tree From Documentation.}
The process for deriving message syntax tree from software documentation can vary among different products and software components. 

With all these information, we can derive the format of a message text as a sequence of different types of elements defined before, which can be further converted to a message syntax tree. For messages with unique IDs, it is easy to associate its syntax tree with its log instance from execution. However, if a message does not have a unique ID for its log instance, we cannot do reverse lookup from log instance to its message documentation. One solution is to a derive a unique signature for these messages from its text structure, keywords, parameter properties etc, that can be also inferred from their log instances. We use the signature of a log message as the index to find its message description and syntax tree. 
Because all the documentations are published online in the forms of webpages and pdf, the general workflow consists of 1). scraping data from documentation media; 2). extracting message IDs and descriptions; 3). building a syntax tree for each message, indexed by message IDs. 
Along with the syntax tree, we may also extract the explanations for a message and the semantic meaning of every parameter in a message from the documentation.

\subsection{Syntax Tree based Message Parsing and Parameter Value Extraction}
Given an instance of a log message, we look up its corresponding log syntax tree by its message ID. After the tokenization of the message text, the parsing can be simply done using depth-first search (DFS) through the tree, which is more efficient and general than using regular expression. The syntax tree based message parsing produces a mapping between variant parts in the log to their corresponding named parameters, referred to as parameter value extraction.    

Using the first log in our previous example as input, DFS will match the tokens \emph{vol1} \& \emph{vol2} with parameter node \emph{volser1} and \emph{volser2}. It results in the correct mapping between variant parts of log messages with the parameter names given in the description for messages in the documentation. The extracted parameter values can be connected with their semantic meanings, which could be very important for log analytic and anomaly detection.

\noindent\textbf{Anchor based Syntax Tree Searching.} In most cases the parameter value is a single token such as status code, system name, ip address, time, etc. However, there are many cases where parameter values consist of
multiple tokens and vary in length. Moreover, the log message and documentation can be noisy. The message text in the documentation can have additional punctuation, conjunctions etc. Any keyword mismatch will fail the search on the syntax tree. To address this problem, we first identify all the sequences consisting of consecutive keywords (of length $>=2$) from the children of the root node. we call them anchors. Then, we search these anchors in a log message to be parsed, divide the text to different parts, each of which is matched with a part of the syntax tree decided by anchors. In Fig 1., consecutive keywords ('OF' ,'VOLUME'), ('TERMINATED', 'ERROR'), ('VTOC', 'COPY', 'DATA', 'SET ') are three anchors. The text of example 1 is divided to multiple ones by these anchors as ('RECOVER'), (' vol1'),  ('ALLOCATING'), ('ON', 'VOLUME', 'vol2). Each is matched with a part of sibling trees between these anchor nodes. For instance, ('RECOVER') is matched with the most left subtree rooted at 'single'.

This approach is able to handle multi-words values in the case that among two anchors there is only one parameter node.  The whole sub-string between two anchors in the log message is treated as the value of that parameter. It also improves the reliability of parsing against noisy data, since the divide-and-conquer approach based on anchor effectively isolates the noisy parts in the log message and the syntax tree from documentation. We can also apply anchor based syntax tree searching recursively to the sub-tree matching. However, because of the cost of finding and matching anchors, it may increase the computation cost. Our observation is that most log messages have a shallow syntax tree, thus for them it is good enough to only find the anchors on the first layer of nodes.  

\subsection{Value Dictionary based Parameter Identification}
Using syntax tree based parsing can identify the parameter values in log messages, and map them to the parameter name given in the documentation. However, it is usually the case that not all log messages have documentation and/or unique message IDs, especially for today's distributed system that runs many different applications and open-source software. The users can also define their own log messages, which are not in the documentation from product vendors. The documentation can also be outdated, which lacks the description of new log messages and/or new parameters added to old log messages. To identify the parameters and their semantics for undocumented ones, we maintain a dictionary of parameter values extracted from documented log messages with their syntax trees. In this \emph{value dictionary}, for each value, we keep a list of parameter names that has been matched with that value. The reason we have a list of names is that the same parameter can have different placeholder names for different messages in the documentation. In addition, for each name, we maintain a count for the number of unique message IDs we've seen that use that name.

For each log message, if it has the syntax tree built from documentation, tree based parsing is used to identify parameters and their values are stored into the value dictionary. If a value already exists in the dictionary, the parameter name list for this value gets updated with new identified parameter name. For the message in example 1, the value dictionary is \{'vol1': [(volser1, count=1], 'vol2': [volser2, count=1]\}.

If the log message does not have the syntax tree, token-based value matching is performed. For each token in the message text, the value dictionary is checked to see if the token is a value seen before or any value begins with that token. If it is, the token (with the following tokens if they together fully matched with that value) is annotated with the name with the largest count in the name list of that value, which could be the most representative name.

Note that different parameters having distinct semantic meanings may share the same value, especially for simple numerical values such as return code, counter value etc. To avoid annotating a value with a wrong parameter, we only keep complex values in the value dictionary for parameter identification, with constraints such as satisfying minimum string length requirement, and no simple numeric values.   

Additionally, for a log message with the syntax tree, after tree-based parsing and parameter name annotation, we can apply the value dictionary based parameter identification for the rest of tokens, in case of missing parameters from the document and the syntax tree, which could be caused by out-of-date or error of documentation. 

% \begin{figure*}
% \centerline{\includegraphics[height=1in, width=5in]{pics/syntaxtree.pdf}}
% \caption{The syntax tree of message ARC0704E}
% \label{fig2}
% \end{figure*}

\subsection{Semantic-aware Template Mining with Value Name Annotation}
In real log data, not all messages or parameters are documented. We cannot solely rely on the syntax tree derived from the documentation for template mining. We use the results of syntax tree based parsing and parameter annotation as an intermediate step, which produces approximate templates from raw logs by replacing the identified values with corresponding parameter placeholders. For example, the approximate template can be "RECOVER OF VOLUME $<$volser1$>$ TERMINATED AT STEP $xx$:$1$" where $<$volser1$>$ is parameter holder with name $volser1$. The approximate templates are further feed into data-driven template mining algorithms such as existing clustering and prefix-tree based parsing to produce final templates that identifies the last step parameter with value $xx$:$1$.

For either clustering or prefix-tree matching method based template mining, a fundamental step is to determine the similarity or measure the distance between two tokens.
With our approximate templates where a token can be a parameter placeholder, the only exception we need to take care of is about how to measure the similarity (for clustering) or decide it is a match (for prefix tree) between a token and a named parameter placeholder. We propose a parameter signature based method to address this problem.

For a parameter, with the values extracted by the syntax tree based parsing, we create a profile for all the values it have hold before, as the parameter's signature. Each time a new value is seen, its signature is updated. To compute the signature, each value is mapped to a profile represented by a feature vector. It consists of the length of the value, and the number of different types of characters within consecutive fixed-size windows over the value string. As an example, given the value "A01-vol1" of $<$volser1$>$, we count the numbers of digits, letters, and other characters in each 4-size window [A01-] and[vol1], which are (1,2,1) and (1,3,0). Combining with the length, we have (8,1,2,1,1,3,0) as the value's profile. A parameter's signature is computed as the average of profile vectors of all distinct values seen by that parameter.

To measure the similarity between a token and a parameter, we compute cosine similarity between token's value profile and the parameter's signature. For similarity between two parameters, we use cosine similarity between their signatures. The cosine similarity can be directly used as distance metric for text clustering. While for prefix-tree matching, we use a threshold such that if the cosine similarity is larger than it, we regard that the token matches with the parameter node in the tree during the traversal search.

\section{EXPERIMENT AND DEMONSTRATION}\label{sec:exp}
Our experiment and demonstration uses system logs collected from IBM mainframe running Z/OS operating systems. We collect MVS message documentation from the pages~\cite{ibmmvs}. We use xml parser to parse the documentation files and extract all documented messages with unique IDs. All the syntax trees are stored in json files, indexed by message IDs. We choose Drain3~\cite{drain3} as our template mining algorithm to process approximate templates. It is an open-source python package that implements Drain~\cite{He2017} algorithm for online log template mining that have been widely used for IBM cloud and has shown the best performance among other template mining methods. We modified Drain3 with prefix-tree matching customized with the matching between tokens and parameters using our parameter signature based method. In addition, the Drain3 algorithm has a template merging operation. For template A and B, it compares each token in A to the token at the same position of B. If two tokens are different, the token in A is replaced with a parameter string $<$*$>$ otherwise is kept. After merging, A becomes the new template. In our scenario, we define priority between parameters and choose to use higher priority parameters in the new template when merging two parameter tokens. Any named parameters has a higher priority than $<$*$>$, and a parameter with a longer name has a higher priority than the one with a shorter name.   

For our approach, we use the first 10000 system logs in a day only for building initial value dictionary as a boot up for our system. We run our approach and Drain separately to perform template mining on the next 10000 logs. We measure the accuracy in the same way as in~\cite{He2017}, with counting true positive (TP) representing two log messages with the same log event being assigned to the same template; false positive (FP) representing two log messages with different events being assigned to the same template; and a false negative (FN) representing two log messages with the same event being assigned different templates. We compute precision, recall and accuracy and compare our approach with Drain3. The result is shown in Table \ref{tab1}. We can see that our approach achieves better accuracy than Drain, which demonstrates the benefit of documentation knowledge and parameter value annotation based on that.   
\begin{table}[htbp]
\caption{Accuracy of Template Mining}
\begin{center}
\begin{tabular}{|c|c|c|c|}
\hline
& precision & recall & accuracy \\
\hline
Our approach& 1.0 & 0.9993 &  0.9996\\
\hline
Drain&  0.7729 & 0.9850 & 0.8661 \\
\hline
\end{tabular}
\label{tab1}
\end{center}
\end{table}

Our demonstration will consist of two steps:

1) We first give a brief overview to our system framework and functionalities, each corresponding to the techniques introduced in the previous sections.

2) We demonstrate the syntax tree building from the document, to log data parsing. We demonstrate the templates output by our approach and compare them with the output of Drain3 by default.

% Figure Labels: Use 8 point Times New Roman for Figure labels. Use words 
% rather than symbols or abbreviations when writing Figure axis labels to 
% avoid confusing the reader. As an example, write the quantity 
% ``Magnetization'', or ``Magnetization, M'', not just ``M''. If including 
% units in the label, present them within parentheses. Do not label axes only 
% with units. In the example, write ``Magnetization (A/m)'' or ``Magnetization 
% \{A[m(1)]\}'', not just ``A/m''. Do not label axes with a ratio of 
% quantities and units. For example, write ``Temperature (K)'', not 
% ``Temperature/K''.

\bibliographystyle{./IEEEtran}
\bibliography{./IEEEabrv,./reference}
\end{document}